\documentclass[twocolumn,twoside,slac,floatfix]{revtex4}
\usepackage{graphicx}
\usepackage{fancyhdr}
\def\pmc{\mbox{PerfMC}}

\begin{document}

\title{A Monitoring System for the BaBar INFN Computing Cluster}

\author{Moreno Marzolla}
\affiliation{Universit\`a Ca' Foscari di Venezia, 30172 Mestre, Italy/INFN Padova, 35100 Padova, Italy}

\author{Valerio Melloni}
\affiliation{Universit\'a di Ferrara, 44100 Ferrara, Italy}

\begin{abstract}
Monitoring large clusters is a challenging problem. It is necessary to
observe a large quantity of devices with a reasonably short delay
between consecutive observations. The set of monitored devices may
include PCs, network switches, tape libraries and other
equipments. The monitoring activity should not impact the performances
of the system.  In this paper we present \pmc, a monitoring system for
large clusters. \pmc\ is driven by an XML configuration file, and uses
the Simple Network Management Protocol (SNMP) for data collection.
SNMP is a standard protocol implemented by many networked equipments,
so the tool can be used to monitor a wide range of devices.  System
administrators can display informations on the status of each device
by connecting to a WEB server embedded in \pmc.  The WEB server can
produce graphs showing the value of different monitored quantities as
a function of time; it can also produce arbitrary XML pages by
applying XSL Transformations to an internal XML representation of the
cluster's status. XSL Transformations may be used to produce HTML
pages which can be displayed by ordinary WEB browsers.  \pmc\ aims at
being relatively easy to configure and operate, and highly
efficient. It is currently being used to monitor the Italian
Reprocessing farm for the BaBar experiment, which is made of about 200
dual-CPU Linux machines.
\end{abstract}

\maketitle

\thispagestyle{fancy}

\section{INTRODUCTION} 
Large clusters with hundreds or thousands of nodes are very difficult
to manage due to their size and the complexity of the applications
they run. Computing farms are routinely used in the current generation
of High Energy Physics experiments, given the huge amount of data to
be processed. 

An efficient monitoring system can be very helpful for profiling all
components of the cluster. A \emph{monitoring system} is a hardware or
software component able to observe the activity of a system
(see~\cite{jain91}). A monitor can observe the performances of a
system, record statistics, analyze the data and display the
results. Monitors are useful for many reasons, such as analyzing the
resource usage of an application or identifying performance
bottlenecks or usage patterns suggesting better algorithms.  Monitors
can be used to characterize the workload of a system, or to find the
parameters to be used on a simulation of a real system. Moreover, they
can be used to check for, and report hardware problems such as crashed
or unresponsive hosts or broken communication links.

Continuous monitoring of computing clusters is a challenging problem
for different reasons. The size (number of components) of the system
to be monitored is often beyond the scalability limit of many
available tools. These tools are usually tailored for a particular
application; they are often closed products, so the user can't adapt
them to any variation of the requirements.  Moreover, they are usually
hard to configure, and provide an inconvenient user interface.

In this paper we present \pmc\ (Performance Monitoring for Clusters),
a monitoring system for large computing clusters. \pmc\ makes heavy
use of XML technologies (see~\cite{xml}), and uses the Simple Network
Management Protocol (SNMP)~(\cite{stallings99}) to gather data from
any device containing an SNMP agent. Since SNMP is a standard
protocol, implemented by many networked equipments, the tool is very
general and can be used to monitor a wide range of devices. \pmc\
embeds a WEB server which is used to generate time-series graphs from
the collected data. Also, the WEB server can produce arbitrary XML
pages by applying transformations to an internal XML representation of
the cluster's status. In this way it is possible to produce HTML
status pages which can be displayed by ordinary WEB browsers.

The paper is organized as follows. Section~\ref{sec:related} presents
some previous works related to the problem of monitoring computing
clusters. In Section~\ref{sec:designgoals} the design goals of \pmc\
are illustrated and discussed. Section~\ref{sec:implementation}
describes the architecture and implementation of \pmc, and a case
study is presented in Section~\ref{sec:casestudy}. Finally,
Section~\ref{sec:conclusions} reports the conclusions and future
works.

\section{RELATED WORKS \label{sec:related}}
In recent years considerable attention has been devoted to the problem
of monitoring the performances of clusters and distributed systems
(see~\cite{ganglia,buyya00,catania96,king00,mansouri95,saab02,subramanyan00,tierney01,Uthayopas:2001:FSR,
wismuller98}). In~\cite{tierney01} the authors describe an agent-based
monitor targeted primarily to GRID architectures (see~\cite{grid}),
which are wide area distributed systems where components can be
connected to high-latency Wide Area Networks. The monitoring architecture is based
on a producer-consumer paradigm, where individual monitors can
subscribe for particular kinds of events, and receive notifications
only when such events are generated by some producer.

It should be noted that a computational GRID is very different in size
and complexity from a computing cluster, so implementing a monitoring
system on them is different. A computational GRID is usually made of
an heterogeneous collection of computing systems which are
geographically distributed and connected through a WAN. A cluster is
made of homogeneous machines, usually residing in the same room and
connected together with a high speed LAN. These differences play an
important role in defining the requirements for a monitoring
system. For example, a monitor for a GRID should be built with
security features, given that intrinsically insecure WAN links are
used for communications. On the other hand, a cluster can be treated
as a single, powerful machine. It should be protected with respect to
the outside world, but communications among machines in the cluster
can be unencrypted. If the monitoring infrastructure is built inside
the cluster, there is no need to protect its control messages. Also,
LANs are characterized by low latencies, high bandwidths and low
packet loss rates. The size of a cluster is usually orders of
magnitude smaller than that of a large GRID.  For these reasons,
monitoring systems developed for computational GRID environments, such
as that described in~\cite{tierney01}, have been developed with very
different requirements in mind with respect to a cluster monitor. 

In~\cite{puliafito00} it is proposed a monitoring and management
architecture based on the use of mobile agents written in Java. Mobile
agents allow management applications to be moved to the network
devices, instead of moving the data provided by the network devices to
the Management Stations. The approach based on Java mobile agents
obviously requires the network devices to be equipped with a Java
Virtual Machine, so that they are able to accept and execute code
coming from the Management Stations. This is currently only possible
when the monitored elements are general-purpose computers, as other
devices are generally unable to run Java code (or any other user
program).

The Ganglia monitoring system described in~\cite{ganglia} is a
distributed monitoring system for clusters. It requires each node on
the cluster to run a d{\ae}mon called \texttt{gmond}. It collects
values from the local machine and broadcasts these values to all the
other \texttt{gmond} processes running on the cluster. To limit the
network utilization, broadcasts happen only when the changes in the
observed values exceed a given threshold.  \texttt{gmond} processes
can also communicate with generic Ganglia clients by sending an XML
status file over a TCP connection. Ganglia d{\ae}mons do not provide
any facility to log the recorded data themselves, but rely on external
programs to collect statistics, perform management actions when
particular events occur, and display the status of the system.

Most monitoring tools use their own data collection protocol over
TCP/IP links. One exception is SIMONE~\cite{subramanyan00}, which uses
the standard SNMP protocol to build a large-scale, distributed
monitoring system.  Hierarchical monitoring has been employed in other
systems as well, such as the one described
in~\cite{Uthayopas:2001:FSR}. Such hierarchical, tree-based monitoring
systems are particularly effective when the user is mainly interested
in getting aggregate informations on the cluster's status, such as the
average load of all the machines, or the least utilized node of the
cluster. This is because the information can be aggregated at each
intermediate node of the hierarchy, thus avoiding the possible
bottleneck of a single node getting all the data from all
hosts. Unfortunately, this strategy does not help when it is necessary
to continuously record the values of some parameters for every single
machine, for example for producing graphs showing the variation of
interesting quantities over the time.

Supermon, described in~\cite{minnich01}, is a centralized monitoring
system, yet it allows efficient and frequent data collection from the
nodes of a Linux cluster. The Linux kernel has to be patched for the
addition of a new system call which provides status informations. A
server program running on each machine collects these informations and
can pass them to requesting applications using a telnet-based network
protocol. A possible drawback of this approach is the necessity to use
a modified kernel on the monitored machines, and the fact that it is
necessary to modify the implementation of the system call if
additional parameters need to be monitored.

\section{\pmc\ DESIGN GOALS \label{sec:designgoals}}
The BaBar Italian reprocessing farm is in production since summer
2002. The farm, hosted at INFN Padova, is made of about 200 dual CPU
Linux/Intel machines, and includes a tape library with a capacity of
70TB, and 24 TB of disk space. The farm is used to process data
collected at SLAC by the BaBar detector.

We identified a number of requirements for the monitoring system
of the cluster, which are now briefly discussed.

{\bf Intrusion-free.} The monitor must be guaranteed not to have
negative impact on the correctness of the monitored system's
results. It should be noted that this cannot be achieved if the
monitored system strongly depends on hard real-time constraints to
operate correctly. For such systems, even a small overhead induced by
the monitoring activity could affect its results. The BaBar
reprocessing farm does not have such constraints.

{\bf Low overhead.} A desirable property of any monitor is that of
imposing a minimal overhead on the observed system. \pmc\ is a purely
software monitor written in C. It uses a clean design in order to be
as efficient as possible. The overhead on the network and on the
monitored devices is extremely low; more details are given in
Section~\ref{sec:casestudy}.

{\bf Batch operation.} The monitoring system should be able to operate
in batch mode, without any user interaction. At the same time, a
suitable user interface should be provided. \pmc\ uses a standard WEB
interface to communicate with the user. This allows the user to
inspect the monitor from remote locations using any WEB browser. 

{\bf Generality.} The monitor should be able to deal with a wide range
of different networked devices, including network switches, tape
libraries, uninterruptible power supplies and so on. The status of
those devices needs to be monitored as well. We identified the Simple
Network Management Protocol (SNMP) as a suitable candidate for the
remote monitoring of a wide range of devices. More details on SNMP
will be given in Section~\ref{sec:snmp}.

{\bf Easy Configuration.} It was very important that the monitoring
system could be configured easily using a standard, structured
notation. We decided to use XML as the language in which the
configuration file is written. There exist many tools able to
generate, verify and transform XML documents. XML parsing can be done
efficiently, as XML documents must obey strict syntactical rules
(see~\cite{xml} for details on XML). In
Section~\ref{sec:configurationformat} we will give more informations
about the structure of the XML-based configuration file.

{\bf Reasonable scalability.} The BaBar INFN reprocessing farm is
expected to grow as the BaBar detector's luminosity increases; this
means that new machines will be added in the future to cope with the
increased volume of data to be processed. The monitoring tool should
be able to scale at least up to moderate cluster sizes (some hundred
nodes).

\section{\pmc\ ARCHITECTURE AND IMPLEMENTATION \label{sec:implementation}}
\pmc\ is a tool for medium-grained, continuous monitoring of computing
clusters. It is written using the C language and currently operates
under the Linux Operating System, but should be easily portable on any
Unix flavor. \pmc\ can monitor any networked equipment implementing an
SNMP agent.
 
\begin{figure*}[ht]
\centering\scalebox{0.7}{\includegraphics{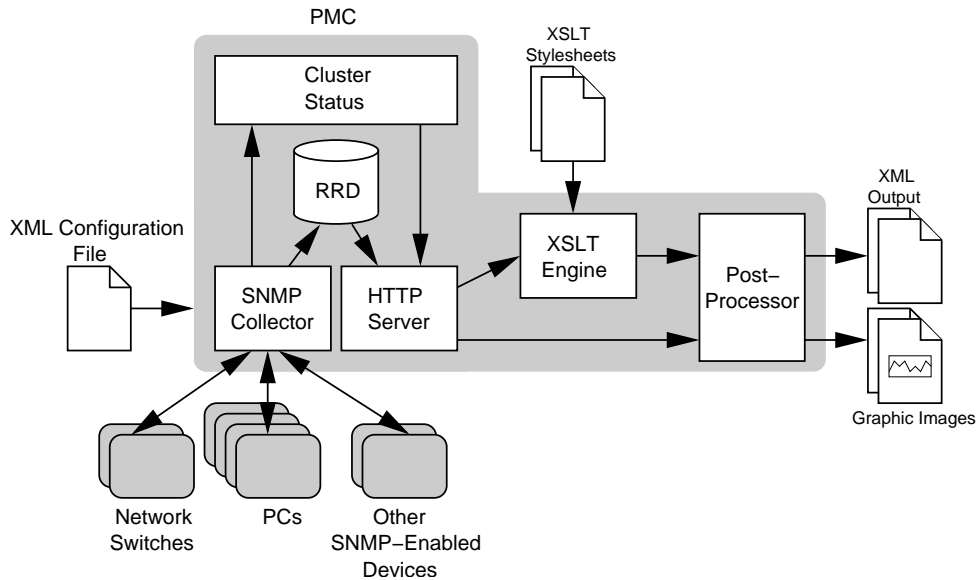}}
\caption{The main components of \pmc.\label{fig:pmc}}
\end{figure*}

\pmc\ is made of two threads: a SNMP collector and
a WEB server, as depicted in Figure~\ref{fig:pmc}.

The SNMP collector thread periodically polls the monitored hosts using
SNMP requests. The list of devices to monitor, and the list of SNMP
variables to request for each one is contained in the XML
configuration file. The collected data are stored on Round Robin
Databases (see~\cite{rrdtools}) on the local disk; at the same time, an
up-to-date ``view'' of the cluster is kept in memory. This view
includes the status of each device (i.e., whether it is responding to
SNMP queries), and the last value for each monitored variable. These
informations are used by the web server thread to produce graphs and
WEB pages. WEB pages are generated by applying user defined XSL
Transformations to an internal XML representation of the cluster
status. It is also possible to define an external postprocessor
through which the generated pages (or graphs) are piped. The SNMP
collector thread and the WEB server thread will be described in more
detail in Sections~\ref{sec:collector} and~\ref{sec:httpserver}
respectively. The format of the configuration file is described in
Section~\ref{sec:configurationformat}.

\subsection{The Simple Network Management Protocol \label{sec:snmp}}
The SNMP architecture has three components:
\begin{itemize}
\item One or more Network Management Stations (NMS), which are responsible
for monitoring and managing other devices;
\item Network Nodes, which may be computing nodes or other equipments;
each node hosts a software component called \emph{SNMP-agent} which
collects local data and answers requests coming from the NMS;
\item A connectionless communication service; SNMP is usually implemented on 
top of UDP.
\end{itemize}

\begin{figure}[t]
\centering\scalebox{0.55}{\includegraphics{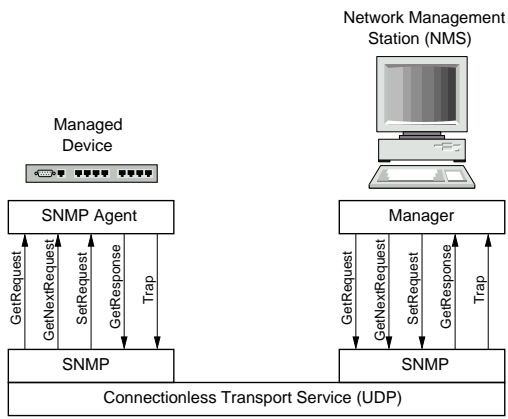}}
\caption{The SNMP Architecture.\label{fig:snmp}}
\end{figure}

The SNMP agent running on each node manages a set of local
\emph{variables}; the management informations pertaining to a
particular class of resources is defined in a Management Information
Base (MIB). The NMS access the MIB by contacting the agent using SNMP
primitives such as \texttt{get} or \texttt{getnext} to read values,
and \texttt{set} to update values. Figure~\ref{fig:snmp} illustrates
the SNMP architecture.

The use of SNMP as a data collection protocol has some drawbacks. The
protocol itself is very simple, and requires each NMS to periodically
poll the other nodes. Polling introduces additional load on the
network, due to the potentially large number of request/response
packets. Also, SNMP agents have a simple structure and usually
communicate only in response to \texttt{get/getnext} requests.  They
don't perform management actions on their own, but require a NMS to
take the decisions about what should be done.

However, SNMP has the advantage of being implemented in virtually
every equipment having a network interface. This includes
workstations, network switches/routers, tape libraries, printers and
Uninterruptible Power Supplies. Many vendors only include SNMP agents
on their devices, and no possibility is offered to use any custom
code; SNMP is thus the only way to interact with these
devices. 

Recent versions of the SNMP protocol have additional features which
solve some of the problems above. In particular, SNMPv2 implements
\emph{bulk requests}, which can significantly reduce the load on the
network by packing several requests into a single datagram. In this
way, the NMS access different MIB variables on the same host by using
a single \texttt{get} request. The responses will be contained in a
single packet as well. Our experience shows that the overhead put on
the network by the SNMP request/response packets is very low. More
details will be given in Section~\ref{sec:casestudy}.

\pmc\ is being tested on a cluster composed of about 200
dual-processor Linux/Intel-based computing nodes, running the SNMP
agent developed by the \textsc{Net-SNMP} (see~\cite{netsnmp}).  

\subsection{The Collector \label{sec:collector}}
The SNMP collector thread is responsible for periodically polling the
various monitored devices. For each device, the user can specify the list
of MIB variables to observe and the frequency of the observations.

In order to improve the efficiency of the collector, multiple hosts
are polled in parallel using non-blocking SNMP requests. The maximum
number of hosts polled in parallel can be defined by the user, the
only limitation being the number of simultaneous opened file
descriptors supported by the underlying Operating System.

The SNMP collector stores the observations into a set of Round Robin
Databases (RRD). A RRD can store time-series data (such as CPU
utilization, network load, machine room temperature) in a compact
way. Data must be entered into a RRD with a certain frequency. Old
data are compacted by means of a consolidation function (any of
Average, Minimum, Maximum and Last), or discarded. For example, the
user may decide to store the average network utilization for the last
week with one observation every 10 seconds, and for the last month
with one observation every minute. The RRDTool package takes care of
compacting observations older than a week by storing the average of
six observations. Data older than one month are discarded.  Round
Robin Databases have constant size, which is determined when they are
created. The RRD library provides the capability to plot the collected
data in various ways. See~\cite{rrdtools} for a complete description
of the RRD package.

The SNMP collector records the status of each machine in the cluster
while receiving the observations. Such informations are kept in
memory. Status informations include, for each host:
\begin{itemize}
\item the last observed value for every SNMP variable;
\item whether the machine is responding to SNMP requests;
\item the list of SNMP error messages generated by the machine.
\end{itemize}
In this way, the collector knows the operational status of each node
in the cluster with a maximum delay equal to the time between
consecutive polls. and can notify the system managers as soon as a
problem arises. 

\subsection{Configuration File Format \label{sec:configurationformat}}
The configuration file for \pmc\ is written in XML. XML documents can
be created by hand using a generic text editor, or using a specialized
XML editor, or automatically generated by an application. The Gnome
XML Library described in~\cite{gnomexml} is used by \pmc\ to parse,
create and transform XML documents. The configuration file conforms to
the structure declared in the \texttt{monitor} Document Type
Declaration (DTD), reported in Figure~\ref{fig:monitordtd}.  XML and
DTD are described in detail in~\cite{xml}.

\begin{figure}[t]
\tiny\begin{verbatim}
<!ELEMENT monitor ( host )+         >
<!ATTLIST monitor
    pmc-num-connections    CDATA #IMPLIED
    pmc-logfile            CDATA #IMPLIED
    pmc-verbosity          CDATA #IMPLIED
    pmc-rrd-dir            CDATA #IMPLIED
    pmc-xslt-dir           CDATA #IMPLIED
    http-html-dir          CDATA #IMPLIED
    http-port              CDATA #IMPLIED
    http-logfile           CDATA #IMPLIED
    http-filter            CDATA #IMPLIED
    http-filter-extensions CDATA #IMPLIED >

<!ELEMENT host (description?, mailto?, miblist, archives, graphs) >
<!ATTLIST host 
    name                ID              #REQUIRED
    ip                  CDATA           #IMPLIED
    polldelay           CDATA           #REQUIRED
    tag                 NMTOKENS        #IMPLIED
    snmpversion         ( 1 | 2c )      "2c"     >

<!ELEMENT description  (#PCDATA)*   >
<!ELEMENT mailto       (#PCDATA)*   >
<!ELEMENT miblist      ( mib )*     >
<!ELEMENT mib          EMPTY        >
<!ATTLIST mib
    id        NMTOKEN                      #REQUIRED
    name      CDATA                        #REQUIRED
    type      ( GAUGE | DERIVE | COUNTER ) "GAUGE"
    community NMTOKEN                      #IMPLIED
    min       CDATA                        #IMPLIED 
    max       CDATA                        #IMPLIED >
  
<!ELEMENT archives     ( rra )*   >
<!ELEMENT rra          EMPTY      >
<!ATTLIST rra
    cf          ( AVERAGE | MIN | MAX | LAST )  "AVERAGE"   
    xff         CDATA                           #IMPLIED
    granularity CDATA                           #REQUIRED
    expire      CDATA                           #REQUIRED  >

<!ELEMENT graphs ( rrdgraph )* >

<!ELEMENT rrdgraph     ( line )+  >
<!ATTLIST rrdgraph
    id          ID     #REQUIRED
    width       CDATA  #IMPLIED
    height      CDATA  #IMPLIED
    seconds     CDATA  #IMPLIED
    title       CDATA  #REQUIRED  >

<!ELEMENT line         (#PCDATA)* >
\end{verbatim}
\caption{The \texttt{monitor} DTD, which defines the structure of the
configuration file for \pmc.\label{fig:monitordtd}}
\end{figure}

The \verb+<monitor>+ tag is the root element of the XML configuration
file. The following optional attributes can be specified:

\begin{description}
\item[pmc-num-connections] The maximum number of concurrent SNMP
connections to use. If this value is set to $N$, then $N$ hosts are
polled in parallel. Default: 50.
\item[pmc-logfile] The log file containing messages generated by
\pmc. Default: do not write any log file.
\item[pmc-verbosity] The level of verbosity of \pmc. Values range from
0 (high verbosity) to 3 (no verbosity). Default: 3 (no verbosity).
\item[pmc-rrd-dir] The directory containing the Round Robin Databases
used to store the observations. Default: current directory.
\item[pmc-xslt-dir] The directory containing the XSL Transformation
files. Default: current directory.
\item[http-html-dir] The directory containing the static HTML files
served by the WEB server. Default: current directory.
\item[http-port] The port on which the WEB server listens to
requests. Default: 8001.
\item[http-logfile] The file containing messages generated by the WEB
server. Default: no log file.
\item[http-filter] The filter (postprocessor) applied to the pages
generated by the WEB server. Default: no filter.
\item[http-filter-extensions] A space-separated list of file
extensions; the filter defined by the \texttt{http-filter} attribute
is applied to every file with a matching extension. Default: empty
extensions list.
\end{description}

The configuration file consists of a sequence of
\verb+<host>...</host>+ blocks, each one containing informations
regarding a specific host to be monitored. The \verb+<host>+ 
tag has the following attributes:

\begin{description}
\item[name] The name of the device. Can be the host name, or any
string used to uniquely identify it.
\item[ip] The IP address of the monitored device.  If not
given, the value of the \texttt{name} attribute is used as the host
name of the device. Default: the value of the \texttt{name} attribute.
\item[polldelay] The delay in seconds between two consecutive
observations of this device.
\item[tag] This attribute may contain any sequence of strings used to
characterize this device. XSL transformations can perform selections
based on the value of this attribute. Default: none.
\item[snmpversion] The version of the SNMP protocol supported by the
device. Recognized values are ``1'' for SNMPv1, and ``2c'' for
SNMPv2. Default: ``2c''.
\end{description}

For each device, the user must provide the list of SNMP variables to
poll and the list of graphs which can be generated for that device,
along with the RRD library commands used to produce the graphs.  SNMP
variables are described by \verb+<mib>...</mib>+ tags.  The following
attributes can be specified:

\begin{description}
\item[id] An identifier for the monitored quantity.  The SNMP
variables of a device needs to be uniquely identified by the value of
this attribute. Variables belonging to different devices may have the
same id.
\item[name] The name of the variable in dotted-decimal
(e.g. ``.1.3.2.4.3.5.4'') or dotted-string (e.g. ``system.sysUpTime'')
notation.
\item[type] Three different kind of SNMP variables are
supported. ``GAUGE'' denotes a variable holding the absolute value of
a quantity (e.g., a temperature).  ``DERIVE'' denotes a variable
holding a value over time (e.g., the number of bytes per second
transmitted over a network interface). ``COUNTER'' denotes a variable
having a value which never decreases (e.g., the total number of bytes
send by a network interface). Default: ``GAUGE''.
\item[community] The name of the SNMP community to which the
variable belongs. Default: ``public''.
\item[min] The minimum value the variable can
assume. Default: none.
\item[max] The maximum value the variable can
assume. Default: none.
\end{description}

\pmc\ creates a RRD for each monitored device. The layout of the RRD
can be specified inside the \verb+<archives>...</archives>+
block. Each RRD is made of a number of Round Robin Archives (RRA),
each described by a \verb+<rra>...</rra>+ tag. Details about the
various options can be found in~\cite{rrdtools}. The attributes are:

\begin{description}
\item[cf] The consolidation function to use. Default: ``AVERAGE''.
\item[xff] The XForm factor, i.e., the fraction of values which must
be inserted into an interval. Default: 0.8;
\item[granularity] The width in seconds of each interval.
\item[expire] The total length in seconds of the RRA. Data older than
this value are discarded.
\end{description}

Finally, \pmc\ is able to create graphs from the recorded data using
the functions provided by the RRDTool package. Each graph is defined
in a \verb+<rrdgraph>...</rrdgraph>+ block, and is characterized by
the following attributes:

\begin{description}
\item[id] The unique identifier of the graph. Graphs from different
devices can have the same id.
\item[width] The width in pixels of the graph. Default: 400.
\item[height] The height in pixels of the graph. Default: 180
\item[seconds] The starting point of the graph expressed in seconds
from the current time, or using the more readable notation supported
by the RRDTool package. Default: ``-3h'' (three hours ago).
\item[title] (none) The title of the graph.
\end{description}

The body of the graphs contain the list of instructions passed
directly to the RRDTool library to produce the graph.

\subsection{The WEB Server \label{sec:httpserver}}
\pmc\ provides a WEB interface through an embedded HTTP server,
implemented using the SWILL library~\cite{swill}. The WEB server has
access to the in-core status informations about the cluster, which is
kept up to date by the SNMP collector thread.  Also, the WEB server
has read-only access to the Round Robin Databases containing the
historical data collected from the cluster. Using the graphing
capabilities provided by the RRD library, the WEB server is able to
dynamically generate plots from the data.

The WEB server can also produce an XML page containing the status of
the whole cluster. As described in Section~\ref{sec:collector}, the status
informations include whether each host is responding to SNMP polls,
the last received values of every polled MIB variable and a list of
error messages reported by the host. The status document includes also
the names of graphs which can be generated for each host in the
cluster. An example of XML status document is reported in 
Figure~\ref{fig:status}.

\begin{figure}[t]
\centering%
\tiny\begin{verbatim}
<?xml version="1.0"?>
<hosts>
  <host name="bbr-farm002" tag="farm1,client" status="OK">
    <mibs>
      <mib id="net2Out" lastUpdated="1018016032">534717280.000000</mib>
      <mib id="net1Out" lastUpdated="1018016032">13811037.000000</mib>
      <mib id="net2In" lastUpdated="1018016032">1741169408.000000</mib>
      <mib id="net1In" lastUpdated="1018016032">13811037.000000</mib>
      <mib id="availSwap" lastUpdated="1018016032">530104.000000</mib>
      <mib id="totalSwap" lastUpdated="1018016032">530104.000000</mib>
      <mib id="totalMem" lastUpdated="1018016032">261724.000000</mib>
      <mib id="cachedMem" lastUpdated="1018016032">35376.000000</mib>
      <mib id="bufferMem" lastUpdated="1018016032">14600.000000</mib>
      <mib id="sharedMem" lastUpdated="1018016032">0.000000</mib>
      <mib id="freeMem" lastUpdated="1018016032">97824.000000</mib>
    </mibs>
    <graphs>
      <graph id="hourly.png" title="Hourly data"/>
    </graphs>
    <notifications>
      <msg ts="1017937775.90771 18:29:35.090771" severity="CRITICAL">Timeout</msg>
    </notifications>
  </host>
<hosts>
\end{verbatim}
\caption{Example of XML status document for a single host. \label{fig:status}}
\end{figure}

The WEB server can apply to the XML status document XSL
Transformations (XSLT) defined by the user. The resulting document is
sent to the user. An XSL Transformation could produce, for example, an
HTML page from the XML status file, so that WEB browsers can display
it. The reader is referred to~\cite{xslt} for details about the XSLT
language.

The user can define in the configuration file an optional
postprocessor, through which every page generated by the WEB server
will be filtered. The command to execute is specified with the
\texttt{http-filter} attribute of the \texttt{monitor} element. The
postprocessor can be any program which accepts input through its
standard input and sends the result to standard output. The user can
restrict the application of the postprocessor to pages with a certain
extension (e.g., only pages with \texttt{.html} or \texttt{.php}
extension). The list of file extensions can be specified with the
\texttt{http-filter-extensions} attribute of the \texttt{monitor}
element.

The WEB server recognizes the following types of Uniform Resource
Identifiers (URI) specified in a HTTP GET request:

{\bf \emph{/$<$hostname$>$/$<$graphname$>$[?querystring]}} Returns the graph
\emph{$<$graphname$>$} for the machine whose identifier is
\emph{$<$hostname$>$}. It is possible to specify a query string for
specifying some parameters of the generated graphs. Recognized query
variables are: {\bf \emph{width}}, for the image width in pixels; {\bf
\emph{height}}, for the image height in pixels; {\bf \emph{start}},
for the starting time of the data plotted on the graph, expressed in
any notation recognized by the RRDTool. For those query variables
which are not specified, the default value specified in the
configuration file for that graph is assumed.  

For example, the graph whose id is ``cpu.png'' for machine
``localhost'', having width of 320 pixels, height of 200 pixels and
showing data starting from 3 hours ago, can be requested with an URI
like this: \url{/localhost/cpu.png?width=320&height=200&start=-3h}

{\bf \emph{/status.html?applyTransform=$<$XSLT-file$>$}} Applies the
XSL Transformation specified in \emph{$<$XSLT-file$>$} to the XML
document describing the whole cluster status. The result is returned
to the user;

{\bf
\emph{/$<$hostname$>$/status.html?~applyTransform~=~$<$XSLTfile$>$}}
Applies the XSL Transformation specified in \emph{$<$XSLT-file$>$} to
the XML document describing the status of the single host
\emph{$<$hostname$>$}. Returns the result to the user.

Generating XML files by applying user-defined transformations to the
status document is particularly useful.  The user can customize the
appearance of the generated HTML pages by writing a set of XSL
Transformations. XSLT is a very powerful language and can be used to
perform any transformations on the XML status document.  Examples of
HTML pages which can be produced are shown in the next section.  

XML is a widely accepted standard, so it qualifies as a suitable mean
for interchanging informations between \pmc\ and other programs. XSL
Transformations can be used to restructure and filter the data
returned by \pmc\ before passing them to an application.

\section{A CASE STUDY \label{sec:casestudy}}
\pmc\ is currently being used to monitor a Linux cluster used for High
Energy Physics applications. The cluster is hosted at INFN Padova,
Italy and is used to process very high volumes of data using
CPU-intensive batch applications. It is made of about 200 dual
processor, Linux/Intel machines, partitioned in about 150 clients and
50 servers. Both classes of machines use 1.26 GHz Pentium III
processors and have 1GB of RAM.  Client machines have two fast
Ethernet controllers (at the moment just one Ethernet board is
used). Server machines have gigabit Ethernet controllers and 1TB of
local disk space as EIDE Raid arrays. The cluster is interconnected
using a high-performance network switch, and is attached to a tape
library.

For each machine in the cluster, the value of the quantities listed in
Table~\ref{tab:mibs} are monitored every 30 seconds. Note that the
stock Net-SNMP agent under Linux does not provide informations on disk
I/O activity. We extended the agent to report these informations by
accessing the \texttt{/proc/stat} file. 

\begin{table}[t]
\centering\begin{tabular}{l|l}
Temp. of the MBoard &
Temp. of the 1st CPU \\
Temp. of the 2nd CPU &
Free Memory \\
Shared Memory &
Buffered Memory \\
Cached Memory &
Total Memory \\
Total Swap &
Available Swap \\
Tot. Disk Blocks read &
Tot. Disk Blocks written \\
Disk 1--5 Blocks read &
Disk 1--6 Blocks written \\
Net 1--4 Bytes in &
Net 1--4 Bytes out \\
\texttt{/tmp} space used &
\texttt{/tmp} space avail \\
\texttt{/var} space used &
\texttt{/var} space avail \\
\texttt{/usr} space used &
\texttt{/usr} space avail \\
Load Average last minute &
Load Average last 5 mins \\
Load Average last 10 mins &
Host ID \\
Host Name &
Host Location \\
Host Uptime & \\
\end{tabular}
\caption{Variables monitored for each host on the test cluster.\label{tab:mibs}}
\end{table}

Every quantity, with the exception of host ID, Name, Location and
Uptime, is stored in a Round Robin Database. For each quantity, the
average and maximum values are recorded in the RRD. Data for the
previous week are kept with the granularity of one observation
(average and maximum) every minute. Data for the previous month are
kept with the granularity of one observation every hour. Data for the
previous year are kept with the granularity of one observation every
day. Finally, data more than one year old are discarded. The total
size of each RRD is about 8 MB, and there is one RRD for each
monitored host.

SNMP bulk get requests are used to get All the monitored variables for
each machine. Network statistics collected with the
\texttt{tcpdump(8)} utility show that the average size of an SNMP
request is, in our case, about $800$ bytes, while the average size of
an SNMP response is about $1000$ bytes. This gives an average network
utilization of approximately 11KB/s due to the monitoring
activity. The CPU overhead on the machines caused by the monitoring is
negligible.  The machine on which \pmc\ runs is a dual Pentium III
running at 1.26 GHz with 1GB or RAM. \pmc\ has a very low (less than
5\%) CPU utilization.

\begin{figure}[t]
\centering%
\scalebox{0.24}{\includegraphics{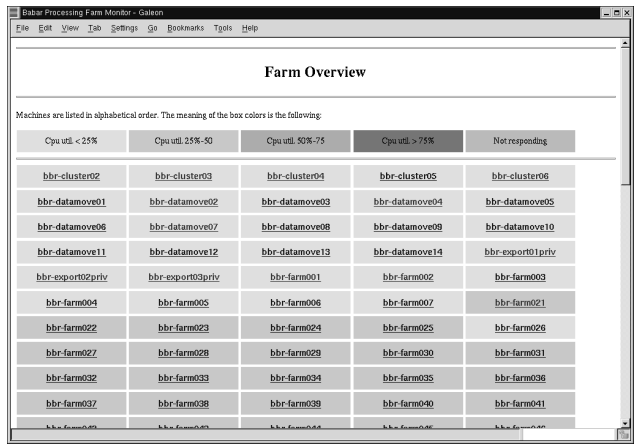}}
\caption{General cluster overview.\label{fig:html1}}
\end{figure}

\begin{figure}[t]
\centering%
\scalebox{0.24}{\includegraphics{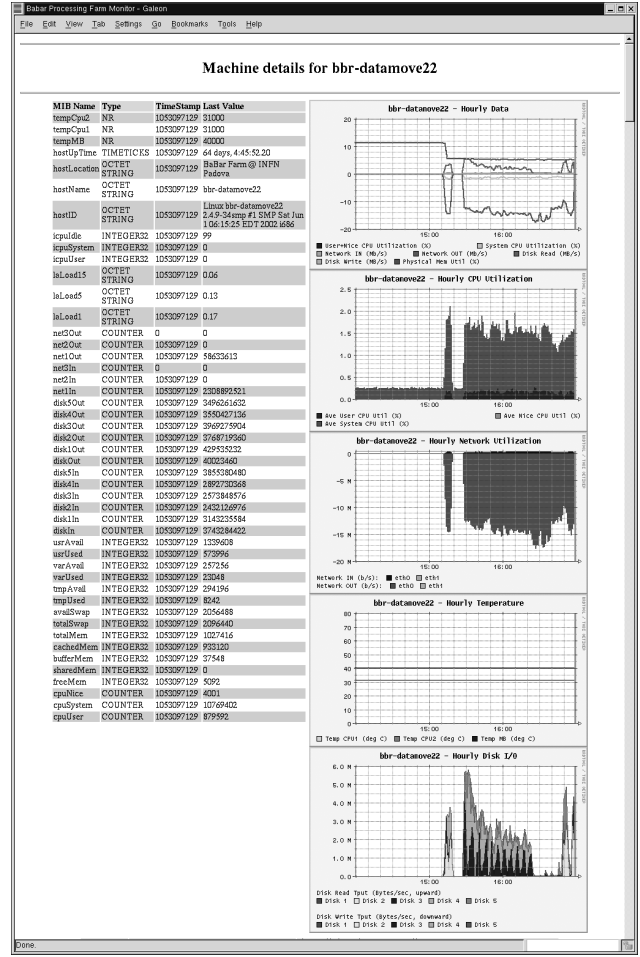}}
\caption{Detailed view for a specific host.\label{fig:html2}}
\end{figure}

Figures~\ref{fig:html1} and~\ref{fig:html2} show two HTML pages
generated by applying two different transformations to the same XML
status document. Figure~\ref{fig:html1} shows a page containing part
of the list of all machines in the BaBar INFN farm, with different
colors indicating the CPU load of each machine. Figure~\ref{fig:html2}
shows a more detailed view of a single machine, with the latest
collected values of all the SNMP variables and some graphs.

\section{CONCLUSIONS AND FUTURE WORK \label{sec:conclusions}}
In this paper we have illustrated the architecture of a monitoring
system for large computing clusters. A prototype written using the C
language, has been implemented and is being used to monitor a cluster
with 200 dual-processor Linux/Intel machines. The monitoring system
uses asynchronous (non-blocking) parallel SNMP bulk requests to
collect status informations from a wide variety of networked devices,
and incorporates a WEB server which can generate graphs from the
collected data. The WEB server can also produce an XML encoding of
the current cluster status, to which an XSL Transformation can
optionally be applied.

We believe that the most essential goals among those stated in
Section~\ref{sec:designgoals} have been satisfied. The system is not
intrusive in that all it needs is an SNMP agent running on each
monitored device. At low polling rates (one observation every $\approx
10$ seconds) the overhead on the network and on the observed devices
is very low.  We do not recommend the use of \pmc\ if higher,
sub-second polling rates are required; in those cases a more
specialized profiling system such as Supermon, should be preferable.

The scalability of \pmc\ has been obtained by a clean design and an
efficient implementation. Polling many hosts in parallel is a very
trivial idea which indeed helped very much.  The use of SNMPv2 bulk
operations allowed to get the values of many variables from a single
machine with just a request/response pair of packet, reducing the load
on the network. Also, SNMP is a standard protocol and is implemented
in virtually every device. The freely available implementation by the
\textsc{Net-SNMP} project allows the user to extend the list of
standard SNMP MIBs without the need to modify the agent's code. In
this way it is possible to monitor everything one could be interested
in. We are currently using this feature to monitor the status and
progress informations from the processes running on our cluster.

Finally, the heavy use of XML as the format for the configuration
file, and that of XSLT to transform the status informations in
arbitrary ways proved to be a good idea. XSLT transformations are used
at the moment to produce a set of HTML pages showing in different ways
the status of our cluster. 

At the moment the prototype does not implement the alarm system. An
alarm system is obviously needed to notify the system administrators
in case of failures, so we are currently working on it. The alarm
system will likely be implemented by listing in the configuration file
a set of thresholds for each MIB variable. If a threshold is crossed
in the specified direction, an alarm will be triggered. An additional
alarm will be associated with each machine, and will be triggered if
the machine does not reply to SNMP queries. Such a threshold-based
alarm system is exactly the same implemented by the RMON
protocol~\cite{stallings99}, so alarms can be triggered directly by
SNMP agents implementing RMON specifications.

The prototype implementation is performing well on our cluster, and no
scalability limit has been encountered so far. However, it is obvious
that a centralized monitoring system, even the most efficient one,
cannot scale forever. In particular, we identified the updating of the
Round Robin Databases as the most likely candidate bottleneck. As a
first solution, we are currently trying to identify possible sources
of inefficiencies in the RRDTool package. As a more long-term fix, we
are considering the idea of partitioning the whole cluster among
different monitors, each one observing a subset of the system. This
would alleviate the scalability problem, as arbitrarily large clusters
can be monitored by simply adding more monitors running on different
machines. On top of these monitors, it is possible to build a
hierarchy of \emph{monitoring proxies} which will be used to fetch and
consolidate the informations collected from the nodes behind them. The
top (root) node will present a global view of the system to the user,
or redirect user's requests to the monitor responsible for observing
the requested resource.  Fault-tolerance can be implemented by means
of standard techniques, such as electing a substitute when one of the
monitors crashes.  The user interface based on HTTP and XML was
developed because it could also be used to exchange informations among
monitors.  The current \pmc\ implementation can be extended to cope
with a hierarchical monitoring infrastructure. What is required is the
addition of a suitable client HTTP interface which can be used to
contact the WEB server embedded in other monitors. In the same way it
is possible to add other kinds of interfaces, such as a Lightweight
Directory Access Protocol (LDAP) interface, or a text-only user
interface.

\begin{acknowledgments}
This work has been partially supported by the Istituto Nazionale di
Fisica Nucleare, Padova. The authors thank Marco Cappellini who
contributed many improvements to \pmc, especially related to the user
interface, and Roberto Stroili for providing comments and suggestions.
\end{acknowledgments}

\bibliography{marzolla-chep03.bib} 

\end{document}